\documentclass[12pt]{article}

\newcommand{\dfr}{\widehat{d}} %fract deriv.
\usepackage[colorlinks=true, pdfstartview=FitV, linkcolor=blue,  citecolor=blue, urlcolor=blue]{hyperref}  %liens hypertexte

\begin{document}
%
% Petites macros:
%
\def\ov{\over}
\def\l{\left}
\def\r{\right}
\def\beq{\begin{equation}}
\def\eeq{\end{equation}}

\def\d{\partial}
% 
%\draft
%\preprint{no. de preprint}

%RŽorienter relativitŽ d'Žchelle-->OK

%Inclure antidiffusion-->OK

%sŽparer pression --> superfluide: chapitre dŽdiŽ, NLSE

%RŽorganiser suivant les 3 reprŽsentations

% titres possibles: Generalized quantum potentials in scale relativity,Representations of the motion equations in the theory of scale relativity

\title{\bf Generalized quantum potentials in scale relativity}
\author{{\bf Laurent Nottale} 
\\
LUTH, CNRS, Observatoire de Paris-Meudon, \\
5 place Jules Janssen, 92195 Meudon Cedex, France \\
e-mail: laurent.nottale@obspm.fr}

\date{\today}
%preprint 
\maketitle

\begin{abstract}
We first recall that the system of fluid mechanics equations (Euler and continuity) that describes a fluid in irrotational motion subjected to a generalized quantum potential (in which the constant is no longer reduced to the standard quantum constant $\hbar$) is equivalent to a generalized Schr\"odinger equation. 
Then we show that, even in the case of the presence of vorticity, it is also possible to obtain, for a large class of systems, a Schr\"odinger-like equation of the vectorial field type from the continuity and Euler equations including a quantum potential. 
The same kind of transformation also applies to a classical charged fluid subjected to an electromagnetic field and to an additional potential having the form of a quantum potential. Such a fluid can therefore be described by an equation of the Ginzburg-Landau type, and is expected to show some superconducting-like properties.
Moreover, a Schr\"odinger form can be obtained for a fluctuating rotational motion of a solid. In this case the mass is replaced by the tensor of inertia, and a generalized form of the quantum potential is derived.
We finally reconsider the case of a standard diffusion process, and we show that, after a change of variable, the diffusion equation can also be given the form of a continuity and Euler system including an additional potential energy. Since this potential is exactly the opposite of a quantum potential, the quantum behavior may be considered, in this context, as an anti-diffusion.
\end{abstract}

%\pacs{PACS numbers: 03.65.Pm, 12.90.+b}
% 03.65.Pm = Relativistic wave equations
% 12.90.+b = Miscellaneous theoretical ideas and models 

%%%%%%%%%%%%
\section{Introduction}  %
%%%%%%%%%%%%
\label{s:intro}

In the theory of scale relativity, three equivalent representations of the equations of motion can be given.

The first is a geodesic form, written in terms of a covariant derivative that includes in its very construction the effects of a nondifferentiable and fractal geometry of space-time \cite{LN93,LN96,CN04,CN06}. For example, in the case of a mere fractal space (i.e., when the  time coordinate itself is not fractal), one finds that the covariant derivative reads ${\dfr}/{dt} ={\partial}/{\partial t}+ {\cal V}. \nabla - 
i {\cal D} \Delta$ \cite{LN93}, where the velocity field ${\cal V}$ is complex as a consequence of a fundamental two-valuedness finding its origin in nondifferentiability. The coefficient ${\cal D}$ is a fundamental parameter which is constant for the system under consideration (the product $2m{\cal D}$ generalizes Planck's constant $\hbar$). The equation of geodesics then takes the simple Galileo inertial-like form of a motion equation in the absence of any force,
\beq
\frac{\dfr  \,{\cal V}}{dt} =0.
\eeq
More generally, by including an external potential $\phi$, one obtains the form of Newton's fundamental equation of dynamics, $m  \,{\dfr} {\cal V}/{dt}+\nabla \phi=0$.

The second representation is a quantum-type, Schr\"odinger form of the equations. Namely, one introduces a wave function $\psi=e^{i{\cal S}/2 m {\cal D}}$, where $\cal S$ is the action, which is now complex since the velocity is itself complex, and which is linked to the velocity field by the relation $ {\cal V}= -2i {\cal D} \nabla \ln \psi$. After performing this change of variable, the equation of dynamics can be integrated under the form of a Schr\"odinger-like equation \cite{LN93},
 \beq
{\cal D}^2 \, \Delta {\psi} + i {\cal D}\,  \frac{\partial{\psi}}{\partial t} - 
\frac{\phi}{2m}{\psi} = 0.
\eeq
One recovers the standard Schr\"odinger equation in the special case ${\cal D}= \hbar/2m$, but the whole mathematical structure of the theory (including the fundamental properties of the solutions) is preserved for a more general value of ${\cal D}$, which can take in particular a macroscopic value \cite[Chap.~7]{LN93}, \cite{LN97,NSL00,DRN03}.

The third representation is a fluid-like system of equations of the Euler and continuity type, including a `quantum potential'. This is a mixed representation with regard to the previous ones, written in terms of the real part $V$ of the complex velocity $\cal V$ and of the squared modulus $P = | \psi |^2$ of the wave function. The real and imaginary parts of the Schr\"odinger equation respectively yield a Euler-like equation (after differentiation) and a continuity equation,
\begin{equation}
 \l(\frac{\partial}{\partial t} + V \cdot \nabla\r) V  = -\nabla \l(\frac{\phi+Q}{m}\r), \;\;\;\;\; \frac{\partial P}{\partial t} + {\rm div}(P V) = 0,
\end{equation}
where an additional  potential energy $Q$ has emerged,  that reads
\begin{equation}
\label{Q1}
Q =-2m{\cal D}^2 \frac{\Delta \sqrt{P}}{\sqrt{P}}. 
\end{equation}
This form of the quantum-mechanical equation has been known for long \cite{Madelung,Bohm}, but 
there are essential differences in the new scale relativistic approach with respect to these earlier studies: (i) In standard quantum mechanics the wave function, the Schr\"odinger equation and the Born postulate are set as axioms of the theory. The variable $V$ is constructed from them, then its meaning as a velocity field must be a posteriori interpreted, without any knowledge about its nature. As a consequence, the meaning and origin of the quantum potential remains unclear. In scale relativity the situation is fundamentally different:  the main axioms of quantum mechanics can be derived from more profound first principles \cite{LN93,LN96,CN04,CN06,LN06,NC07} (namely, the principle of relativity itself, once it is extended to scale transformations of the reference system), the velocity field $V$ is introduced from the very beginning as that of the geodesics fluid, and the emergence of the quantum potential can be understood as a mere manifestation of the fractal geometry of space-[time] (in analogy with general relativity, where the Newton potential is understood as a manifestation of the curved geometry of space-time). (ii) The amplitude of the quantum potential can now be macroscopic, while in standard quantum mechanics it is constrained to be proportional to the microscopic constant $\hbar^2$.

In the present paper, we study the reversibility of the transformations between these various representations and we generalize them to more complicated situations. As we shall see, still generalized forms of the quantum potential can be found, and the application of such a quantum potential to classical Euler and continuity equations allows one to integrate them into Schr\"odinger-type equations. In such a framework, it should be clear from now on that by `Schr\"odinger equation' we no longer exclusively mean the particular equation of atomic and molecular quantum physics, but more generally a generic and universal form taken by the equations of dynamics under some general conditions (nondifferentiability and fractality, or presence of a quantum-like potential).

In these derivations, the quantum potential is no longer founded on the quantum Planck's constant $\hbar$, but on a more general parameter $\cal D$ which can take any macroscopic value and which gives back standard quantum mechanics only in the special case ${\cal D}= \hbar/2m$. Therefore a fluid that would be subjected to such a generalized quantum-like potential is expected to exhibit some macroscopic properties typical of quantum fluids (though certainly not every aspects of a genuine quantum system). It has already been suggested that this could already be the case of some astrophysical systems \cite{LN93,NSL00,DRN03,LN96B,NSG97}, and of some functions, structures and properties typical of living systems \cite{LN04,AN07,NA07}. Moreover, since the Euler and continuity system of equations can also be used as approximation in the description of many other types of physical systems, such as chaotic mechanical systems, $n$ particle dynamics, etc... (see e.g. \cite{LiLi83}), these results may also be relevant in these cases.

%%%%%%%%%%%%
\section{Scale relativity theory}  %
%%%%%%%%%%%%

%%%%%%%%%%%%%%%%%%%%%%%%%%%%%%%%%%%%%%%%%%%%%
\subsection{Geodesics equation in a fractal space}
\label{sec2.1}
%%%%%%%%%%%%%%%%%%%%%%%%%%%%%%%%%%%%%%%%%%%%%

In previous works \cite{LN93,CN04,NC07}, we have given, in the framework of the theory of scale relativity and fractal space-time, a derivation of a generalized Schr\"odinger equation and of a generalized Compton relation. Let us briefly remind the main steps of this derivation.

One can prove  \cite{LN93,LN96,CN04} that the elementary displacements $dX$ on the geodesics of a nondifferentiable fractal space-time can be decomposed as the sum of two terms,
\beq
d_{\pm}X = d_{\pm}x + d_{\pm}\xi,
\label{eq.18}
\eeq
$d_{\pm}Ê\xi$ representing the ``fractal (differentiable) part'' and $d_{\pm}x$, the ``classical (non-differentiable) part'', 
defined as
\beq
d_{\pm}x = v_{\pm} \; dt,
\label{eq.19}
\eeq
\beq
d_{\pm}\xi=\eta_{\pm} \,  \sqrt{2 \cal{D}}\,  dt^{1/2},
\label{eq.20bis}
\eeq
where $\eta$ is a normalized stochastic (or simply fluctuating) variable such that $\langle \eta \rangle=0$ and $\langle \eta^2 \rangle=1$. This expression for the fractal fluctuation $d_{\pm} \xi$ corresponds to the critical fractal dimension $D_F=2$ (see \cite{LN96}). The two-valuedness of these differentials (described by the subscript $\pm$) is a consequence of a breaking of the invariance under infinitesimal time reflexion ($dt \leftrightarrow -dt$) that comes directly from nondifferentiability \cite{CN04}.

Then one combines  the two time derivatives in terms of a complex derivative operator \cite{LN93}
\beq
\frac{\dfr}{dt} = \frac{1}{2} \left( \frac{d_+}{dt} + \frac{d_-}{dt} \right) 
- \frac{i}{2} \left(\frac{d_+}{dt} - \frac{d_-}{dt}\right) \; .
\label{eq.27}
\eeq
Applying this operator to the position vector yields a complex velocity 
\beq
{\cal V} = \frac{\dfr}{dt} x(t) = V -i U = \frac{v_+ + v_-}{2} - i 
\;\frac{v_+ - v_-}{2} \; .
\label{eq.28}
\eeq
This complex time derivative operator has been found to be given by \cite{LN93}
\beq
\frac{\dfr}{dt} = \frac{\partial}{\partial t}+ {\cal V}. \nabla - 
i {\cal D} \Delta \; .
\label{eq.32}
\eeq
It  plays the role of a ``covariant derivative operator'', i.e., of a tool that allows the equation of dynamics in a fractal nondifferentiable space to be given the same form as in a classical differentiable one.

Namely, the classical 
part of the system can be characterized by a Lagrange function 
${\cal L} (x, {\cal V}, t)$, from which an action ${\cal S}$ is defined
\beq
{\cal S} = \int_{t_1}^{t_2} {\cal L} (x, {\cal V}, t) dt,
\label{eq.33}
\eeq
where both the Lagrange function and the action are now complex because the velocity $\cal V$ is itself complex. Using this covariant derivative, one then write the equation of dynamics in a potential $\phi$ under Newton's classical form (although this equation is no longer classical)
\beq
m \frac{\dfr}{dt} {\cal V}= - \nabla \phi .
\label{eq.37}
\eeq
In the case when there is no external field the covariance is 
explicit, since Eq.~(\ref{eq.37}) can be identified with a  geodesics equation that takes the form of Galileo's equation of inertial motion
\beq
\frac{\dfr}{dt} {\cal V} = 0.
\label{eq.37bis}
\eeq
This vacuum form of the motion equation is also obtained when the field can be itself derived from a covariant derivative under a geometric interpretation, as in the case of gravitation in Einstein's theory of general relativity of motion, and now of gauge fields \cite{NCL06}.

In both cases (with and without external field), the complex momentum ${\cal P}$ reads
${\cal P} = m {\cal V} $ \cite{LN93},
so that the complex velocity ${\cal V}$ is 
 the gradient of the complex action,
${\cal V} = \nabla {\cal S}/ m$.

One then introduces a complex function $\bar{\psi}$ which is nothing but another 
expression for the complex action ${\cal S}$,
\beq
\bar{\psi} = e^{i{\cal S}/{\cal S}_{0}}.
\label{eq.40}
\end{equation}
The factor ${\cal S}_{0}$ must be 
introduced for dimensional reasons and has the dimension of an action (namely, it is identified with $\hbar$ in the case of standard quantum mechanics). The $\bar{\psi}$ function 
is therefore related to the complex velocity 
as follows
\beq
{\cal V} = - i \, \frac{{\cal S}_{0}}{m} \, \nabla (\ln \bar{\psi}),
\label{eq.41}
\eeq
so that the fundamental equation of dynamics of Eq.~(\ref{eq.37}) reads
\beq
i {\cal S}_{0} \frac{\dfr}{dt}(\nabla \ln \bar{\psi}) = \nabla \phi.
\label{eq.42}
\eeq
Replacing $\dfr/dt$ by its expression, given by Eq.~(\ref{eq.32}), 
and replacing ${\cal V}$ by its expression in Eq.~(\ref{eq.41}), 
one obtains 
\beq
\nabla   \phi  =   i {\cal S}_{0} \left[ \frac{\partial }{\partial t} \nabla   
\ln\bar{\psi}   - i \left\{  \frac{{\cal S}_{0}}{m} (\nabla   \ln\bar{\psi}  . \nabla   )
(\nabla   \ln\bar{\psi} ) + {\cal D} \Delta (\nabla   \ln\bar{\psi} )\right\}\right] .
\label{eq.44}
\eeq

%*************************************
\subsection{Basic remarkable identity}
\label{ident}
 
Let us prove a general remarkable identity that plays a central role in the mathematical structure of the theory. Start from the following derivation (which is general, including the general relativity case where the indices are raised and lowered using a metric tensor)
\begin{eqnarray}
\partial_{\mu} \partial^{\mu} \ln R +\partial_{\mu} \ln R\; \partial^{\mu} 
\ln R 
&=& \partial_{\mu} \frac{\partial^{\mu} R}{R}+\frac{\partial_{\mu} R}
{R}\frac{\partial^{\mu} R}{R} \nonumber \\
&=& \frac{R \, \partial_{\mu} \partial^{\mu} R - 
\partial_{\mu} R \, \partial^{\mu} R}{R^{2}}+\frac{ \partial_{\mu} R \, 
\partial^{\mu} R}{R^{2}} \nonumber \\
&=& \frac{\partial_{\mu} \partial^{\mu} R}{R} \; . 
\label{eq.46a}
\end{eqnarray}
Returning to the operatorial notation, this identity reads  \cite{LN93}
\beq
\frac{\Delta R}{R} =(\nabla \ln R)^{2} + \Delta \ln R.
\label{eq.45a}
\eeq
By taking its gradient, we obtain
\beq
\nabla \left(\frac{\Delta R}{R}\right)=\nabla \l\{(
\Delta \ln R + \nabla \ln R)^{2}  \r\}.
\label{eq.47a}
\eeq
The first term in the right-hand side of this expression can be transformed, 
using the fact that $\nabla$ and $\Delta$ commute, i.e., $\nabla \Delta =\Delta \nabla$.
The second term can also be transformed thanks to another remarkable identity
\beq
\nabla (\nabla f)^{2}=2 (\nabla f . \nabla) (\nabla f) ,
\label{eq.49a}
\end{equation}
that we apply to $f=\ln R$. We finally obtain \cite{LN93}
\beq
\nabla\left(\frac{\Delta R}{R}\right)=  2 (\nabla \ln R . \nabla )(\nabla \ln R )+ \Delta (\nabla \ln R).
\label{eq.50a}
\eeq
This identity can be still generalized thanks to the fact that $R$ appears only through its logarithm in the right-hand side of the above equation. By replacing in it $R$ by $R^{\alpha}$, we obtain the general remarkable identity
 \beq
\frac{1}{\alpha} \; \nabla\left(\frac{\Delta  R^{\alpha}}{R ^{\alpha}}\right)= 2\alpha (\nabla \ln R . \nabla )(\nabla \ln R) + \Delta (\nabla \ln R),
 \label{remid}
\eeq
in which the numerical coefficient in the linear combination of the right-hand side is no longer constrained to be 2 but can now be any number. More generally, one obtains a gneral relation that reads in tensorial notation, including the case when the indices are raised and lowered by a metric tensor,
\beq
\frac{1}{\alpha} \;\d_k \frac{\partial_{\mu} \partial^{\mu} R^\alpha}{R^\alpha}= 2\alpha \,\partial_{\mu} \ln R\; \partial^{\mu}\d_k \ln R+ \partial_{\mu} \partial^{\mu}\d_k \ln R.
\label{idgen}
\eeq
 
 %***************************************************************
\subsection{Derivation of generalized Schr\"odinger equation and Compton relation}
%******************************************************************************************
 
Equation~(\ref{eq.44}) can be put under the form
 \beq
\nabla   \phi  =   i {\cal S}_{0} \left[ \frac{\partial }{\partial t} \nabla   
\ln \bar{\psi}   - i {\cal D} \left\{  \frac{{\cal S}_{0}}{m {\cal D} } (\nabla   \ln \bar{\psi}   . \nabla   )
(\nabla   \ln \bar{\psi}  ) +\Delta (\nabla   \ln \bar{\psi}  )\right\}\right] .
\label{eq.100}
\eeq
Then using the generalized identity (\ref{remid}) in which we set $\alpha=S_0/2m{\cal D}$, we obtain
\beq
  \frac{{\cal S}_{0}}{m {\cal D} } \, (\nabla   \ln \bar{\psi}   . \nabla   )
(\nabla   \ln \bar{\psi}  ) +\Delta (\nabla   \ln \bar{\psi}  )= \frac{2m{\cal D}}{S_0} \,
 \nabla
 \l(
  \frac{\Delta \bar{\psi} ^{\alpha } }          { \bar{\psi} ^{\alpha}}
  \r).
\eeq
Therefore the right-hand side of equation~(\ref{eq.100}) becomes a gradient whatever the value of $S_0$, namely
\beq
 \nabla \phi = i {\cal S}_{0} \, \nabla \left[ \frac{\partial}{\partial t} \ln  \bar{\psi}  - i  \, \frac{2m{\cal D}^2}{S_0} \,
  \frac{\Delta \bar{\psi}^{\alpha } }          { \bar{\psi}^{ \alpha}}
   \right].
\label{eq.101}
\eeq
Let us now define a new function
\beq
\psi=   {\bar{\psi}^{\alpha}}= {\bar{\psi}^{    ({S_0}/{2m{\cal D}}) }}= e^{i {\cal S}/{2m {\cal D}}}.
\eeq
 
The two functions are related by the equation  $\ln \bar{\psi}= (2m {\cal D} /S_0) \ln{\psi}$, so that equation~(\ref{eq.101}) becomes
 \beq
  \nabla \phi = i \, 2m {\cal D}  \, \nabla \left[\frac{ \partial{\psi }/{\partial t} - i  \, {\cal D} \Delta {\psi}}{\psi}
   \right].
 \eeq
 It can finally, without any further hypothesis,  be integrated under the form of a generalized Schr\"odinger equation
 \beq
{\cal D}^2 \, \Delta {\psi} + i {\cal D}\,  \frac{\partial{\psi}}{\partial t} - 
\frac{\phi}{2m}{\psi} = 0.
 \label{schro}
\eeq
More generally one obtains the same Schr\"odinger equation even when including the fractal (divergent) part of the complex velocity field in the covariant derivative \cite{LN06,LN99} and therefore in the very definition of the wave function. This implies that the nondifferentiability of space can manifest itself in the fractality and nondifferentiability of the wave function. This result agrees with the recent discovery by Berry \cite{MB96} and Hall \cite{MH04} of fractal solutions to quantum mechanical equations.

The constant $S_0$ is now fixed at the value
\beq
S_0=2 m {\cal D}.
\label{compton}
\eeq
In the previous derivations \cite{LN93,CN04}, the  relation $S_0=2 m {\cal D}$ was assumed as a necessary condition for transforming the geodesics equation in a Schr\"odinger-like equation. Here it has been derived without any additional hypothesis. This is an important point, since this relation is nothing but a generalized Compton relation. Indeed, the parameter $\cal D$ is a mere re-expression of a fundamental length scale of the theory, $\lambda_c=2{\cal D}/c$. Now, in the standard quantum case, $S_0$ can be taken to be a fundamental and universal constant, namely, it identifies with the Planck constant, $S_0=\hbar$. In this case, the relation (\ref{compton}) becomes indeed the standard Compton relation, $\lambda_c=\hbar/mc$, and the equation (\ref{schro}) becomes the standard Schr\"odinger equation.

But we stress, in the scale relativity approach, that:

(i) The whole physical and mathematical structure remains self-consistent even for a value of $S_0$ different from $\hbar$, provided it remains a constant for the system under consideration. 

(ii) The Compton length receives in this framework a new interpretation, namely,  it gives the amplitude of the fractal fluctuations. Indeed, this amplitude reads under the form $(d \xi/\lambda_c)^2=\eta^2 (c \,dt/\lambda_c)$, so that the Compton length (and therefore the mass in standard quantum theory) can be defined in a geometric way as
\beq
\lambda_c= \frac{\langle d \xi^2  \rangle}{c \,dt},
\eeq
where $\langle \rangle$ denotes averaging over the stochastic or fluctuating variable $\eta$, whatever may be its probability distribution. 

(iii) The de Broglie length also acquires, in such a framework, a simple geometric interpretation as a classical to fractal transition (in scale space). Indeed, since $dx$ and $dt$ are differential elements of the same order, one may write $d \xi^2= \eta^2 \times  \lambda_c \, c dt$ under the form $d \xi ^2= \eta^2 \times  \lambda dx$. The new length scale introduced in this expression for dimensional reasons therefore reads $\lambda=c \lambda_c/(dx/dt)$, i.e.,
\beq
\lambda= \frac{ c}{v}\times \lambda_c,
\eeq
which generalizes the non relativistic de Broglie scale since this gives $\lambda= \hbar/mv$ when $\lambda_c=\hbar/mc$ in the standard quantum case. The elementary displacements therefore read $dX= dx + d \xi= dx(1+\eta \sqrt{\lambda/dx})$, and they indeed show a transition from a classic, differentiable behavior to a fractal, nondifferentiable (i.e., scale-divergent) behavior when $dx$ becomes smaller than $\lambda$.

This result is an important point as concerns the possible applications of the theory, since it proves that the full mathematical and physical structure (Schr\"odinger equation and Compton-de Broglie relations for the wave function which is a solution of it) is preserved in a scheme more general than the standard quantum one ( which corresponds to the very particular case ${\cal D}=\hbar/2m$). This result therefore involves the possibility to build macroscopic quantum-type systems which are no longer constrained by the Planck constant (for example systems embedded in fractal media which would be scaling on a large range of scales \cite{LN93}) and/or the possibility that such systems do already exist in nature \cite{LN93,LN97,NSL00,DRN03,LN96B,LN04}.

%******************************************************
\section{Schr\"odinger equation in fluids mechanics}
%******************************************************

\subsection{From Schr\"odinger to Euler and continuity equations}
%********************************************************************
\label{sec3.1}

By separating the real and imaginary parts of the generalized Schr\"odinger equation and by using a mixed representation of the motion equations in terms of ($P, V$), instead of ($V,U$) in the geodesics form and  ($P, \theta$) in the Schr\"odinger form, one obtains fluid dynamics-like equations, i.e., a Euler equation and a continuity equation (this is a generalization of the Madelung-Bohm transformation, but whose physical meaning is set from the very beginning instead of being a posteriori interpreted). 

Let us recall explicitly this transformation.  We first come back to the definition of the wave function by making explicit the probability and the phase, namely,
\beq
\psi = \sqrt{P } \times e^{i S/{2m {\cal D}}}.
\eeq
By introducing this form of the wave function in the Schr\"odinger equation~(\ref{schro}) with an exterior scalar potential $\phi$, we obtain
%cahier21,81
\beq
\l\{-\frac{\sqrt{P}}{2m} \l( \frac{\d S}{\d t} +\frac{(\nabla S)^2}{2m}+ {\phi}-2m{\cal D}^2 \frac{\Delta \sqrt{P}}{\sqrt{P}}\r) + i \frac{\cal D}{2\sqrt{P}}\l( \frac{\partial P}{\partial t} + {\rm div}(P \frac{\nabla S}{m})  \r) \r\} e^{i S/2m{\cal D}}=0.
\label{rischro}
\eeq
Now the complex velocity ${\cal V}=V-iU$ being linked to the wave function by the relation ${\cal V}=-2i {\cal D} \nabla \ln \psi$, its real part is therefore given in terms of the phase by the standard classical relation \cite{LN93}
\beq
V=\frac{\nabla S}{m}.
\eeq
We note once again that this fundamental identification is here derived (since $V$ has been defined from the very beginning as the real part of the geodesics mean velocity field), while in the standard Madelung transformation $V$ is defined from the above equation itself, and it is therefore interpreted from it. In the scale relativity / nondifferentiable space-time approach, the velocity field and therefore the wave function from which it derives characterize from the beginning the bundle of potential fractal geodesics. 

By replacing in the above form of the Schr\"odinger equation $\nabla S/m$ by the real velocity field $V$, it reads
\beq
\l\{-\frac{\sqrt{P}}{2m} \l( \frac{\d S}{\d t} +\frac{1}{2}m V^2+ {\phi}+Q\r) + i \frac{\cal D}{2\sqrt{P}}\l( \frac{\partial P}{\partial t} + {\rm div}(P V)  \r) \r\} e^{i S/2m{\cal D}}=0.
\label{rischro2}
\eeq
The real part of this equation is an energy equation,
\beq
E=- \frac{\d S}{\d t}=\frac{1}{2}m V^2+ {\phi}+Q,
\eeq
whose gradient yields an Euler-type equation
\begin{equation}
\label{AA1}
 \l(\frac{\partial}{\partial t} + V \cdot \nabla\r) V  = -\nabla \l(\frac{\phi+Q}{m}\r),
\end{equation}
while the imaginary part is a continuity equation, namely,
\begin{equation}
\label{AA2}
\frac{\partial P}{\partial t} + {\rm div}(P V) = 0.
\end{equation}
But, in the energy and Euler equations, an additional  potential energy $Q$ has emerged,  that writes
\begin{equation}
\label{Q}
Q =-2m{\cal D}^2 \frac{\Delta \sqrt{P}}{\sqrt{P}}. 
\end{equation}
This scalar potential generalizes to a constant $\cal D$ which may be different from $\hbar/2m$ the quantum potential obtained in the Madelung-Bohm transformation.  The potential $Q$ is now understood as a manifestation of the fractal geometry, and the probability density is also interpreted in this framework as arising from the distribution of geodesics, so that the Born postulate is derived \cite{CN04,LN06,NC07}, as can be verified by numerical simulations \cite{RH98,LN07}.

%*********************************
\subsection{Inverse derivation: from Euler to Schr\"odinger equation (pressure-less potential motion)}
%******************************
 It is less well known that the transformation from the generalized Schr\"odinger equation to the Euler and continuity equations with quantum potential is reversible. Indeed, the Euler  and continuity system reads in the pressure-less case
 \begin{equation}
\label{BBB1}
 \l(\frac{\partial}{\partial t} + V \cdot \nabla\r) V  = -\nabla \l({\phi}-2{\cal D}^2 \frac{\Delta \sqrt{\rho}}{\sqrt{\rho}}\r),
\end{equation}
\begin{equation}
\label{BBB2}
\frac{\partial \rho}{\partial t} + {\rm div}(\rho V) = 0.
\end{equation}
Their form is similar to Eqs.~(\ref{AA1}) and (\ref{AA2}), but with the probability density $P$ replaced by the matter density $\rho$ and with $m=1$. Assume, as a first step, that the motion is irrotational (see the following Section~\ref{vorticity} for the account of vorticity and pressure). Then we set
\beq
V=\nabla S.
\eeq
Equation~(\ref{BBB1}) takes the successive forms
\beq
\frac{\d}{\d t}(\nabla S) +\frac{1}{2} \nabla (\nabla S)^2+ \nabla \l({\phi}-2{\cal D}^2 \frac{\Delta \sqrt{\rho}}{\sqrt{\rho}}\r)=0,
\eeq
\beq
\nabla \l( \frac{\d S}{\d t} +\frac{1}{2}(\nabla S)^2+ {\phi}-2{\cal D}^2 \frac{\Delta \sqrt{\rho}}{\sqrt{\rho}}\r)=0,
\eeq
which can be integrated as
\beq
 \frac{\d S}{\d t} +\frac{1}{2}(\nabla S)^2+ {\phi}+C-2{\cal D}^2 \frac{\Delta \sqrt{\rho}}{\sqrt{\rho}}=0,
 \eeq
where $C$ is a constant that can be taken to be zero by a redefinition of the potential energy $\phi$. Let us now combine this equation with the continuity equation as follows:
\beq
\label{CCC}
\l\{-\frac{1}{2} \sqrt{\rho}\l( \frac{\d S}{\d t} +\frac{1}{2}(\nabla S)^2+ {\phi}-2{\cal D}^2 \frac{\Delta \sqrt{\rho}}{\sqrt{\rho}}\r) + i \frac{\cal D}{2\sqrt{\rho}}\l( \frac{\partial \rho}{\partial t} + {\rm div}(\rho \nabla S)  \r) \r\} e^{i S/2{\cal D}}=0.
\eeq
We have therefore recovered the form (\ref{rischro}) of the Schr\"odinger equation (with $m=1$).
Finally we set
\beq
\label{psi}
\psi=\sqrt{{\rho}} \times e^{{i S}/{2 {\cal D}}},
\eeq
and the equation~(\ref{CCC}) is strictly identical to the following generalized Schr\"odinger equation:
\beq
{\cal D}^2 \Delta \psi + i {\cal D} \frac{\partial}{\partial t} \psi - \frac{\phi}{2}\psi = 0.
\eeq
 Given the linearity of the equation obtained, one can normalize the modulus of  $\psi$ by replacing the matter density $\rho$ by a probability density  $P=\rho/M$, where $M$ is the total mass of the fluid in the volume considered. These two representations are equivalent.

The imaginary part of this generalized Sch\"odinger equation amounts to the continuity equation and its real part to the energy equation that reads
\beq
E=- \frac{\d S}{\d t} =\frac{1}{2}mV^2+ {\phi}-2{\cal D}^2 \frac{\Delta \sqrt{\rho}}{\sqrt{\rho}}.
\eeq

%*********************************
\subsection{From Euler to Schr\"odinger: account of pressure}
\label{pressure}
%******************************
Consider now the Euler equations with a pressure term and a quantum potential term: 
\begin{equation}
\label{AAA1}
 \l(\frac{\partial}{\partial t} + V \cdot \nabla\r) V  = -\nabla \phi-\frac{\nabla p}{\rho}+2{\cal D}^2 \,\nabla \l( \frac{\Delta \sqrt{\rho}}{\sqrt{\rho}}\r).
\end{equation}
When ${\nabla p}/{\rho}=\nabla w$ is itself a gradient, which is the case of an isentropic fluid, and, more generally, of every cases when there is an univocal link between pressure and density, e.g., a state equation \cite{Landau6}, its combination with the continuity equation can be still integrated in terms of a Schr\"odinger-type equation \cite{LN97},
\beq
{\cal D}^2 \Delta \psi + i {\cal D} \frac{\partial}{\partial t} \psi - \frac{\phi+w}{2}\psi = 0.
\eeq
Now the pressure term needs to be specified through a state equation, which can be chosen as taking the general form $p=k_p \, \rho^{\gamma}$.

In particular, in the sound approximation, the link between pressure and density writes $p-p_0=c_s^2(\rho-\rho_0)$, where $c_s$ is the sound speed in the fluid, so that $\nabla p/\rho=c_s^2 \, \nabla \ln \rho$. In this case, which corresponds to $\gamma=1$, we obtain the non-linear Schr\"odinger equation
\beq
{\cal D}^2 \Delta \psi + i {\cal D} \frac{\partial}{\partial t} \psi -k_p\, \psi  \ln |\psi|  =\frac{1}{2} \, \phi  \; \psi,
\eeq
with $k_p=c_s^2$. When $\rho-\rho_0 \ll \rho_0$, one may use the additional approximation $c_s^2 \, \nabla \ln \rho \approx (c_s^2 /\rho_0) \nabla \rho$, and the equation obtained takes the form of the non-linear Schr\"odinger equation which is encountered in the study of superfluids and of Bose-Einstein condensates (see e.g. \cite{Nore, Fetter1965}), and which is also similar to the Ginzburg-Landau equation of superconductivity \cite{Landau9} (here in the absence of field),
\beq
{\cal D}^2 \Delta \psi + i {\cal D} \frac{\partial}{\partial t} \psi - \beta |\psi|^2 \, \psi= \frac{1}{2} \, \phi  \; \psi,
\eeq
with $\beta={c_s^2}/{2 \rho_0}$. In the highly compressible case the dominant pressure term is rather of the form  $p \propto \rho^2$, so that $p/\rho \propto \rho= |\psi|^2$, and one still obtains a non-linear Schr\"odinger equation of the same kind \cite{Nore}.

%1.5***********************************************************************************************
\subsection{From Schr\"odinger equation in vectorial field to Euler and continuity equations}
\label{sec2.5}

Let us now consider a more general case. In the previous sections, only a scalar external field was taken into account. We shall now study the decomposition of the Schr\"odinger equation which applies to a system subjected to a vectorial field (such as, e.g., a magnetic field). As we shall now show, it can also be generally decomposed in terms of a Euler-type equation and a continuity-type equation, with the external vectorial field playing a role similar to the rotational part of the velocity field. Thanks to this analogy, this decomposition applies actually to two different cases: (i) quantum fluids subjected to a magnetic field (such as in the Ginzburg-Landau equation of superconductivity); (ii) some fluids with a non-potential velocity field. 

Start from the general form of the Schr\"odinger equation for a spinless particle subjected to a scalar field $\phi$ and to a vectorial field $K_j$ (for example, an electromagnetic field):
\beq
\l\{\frac{1}{2} (-2i {\cal D} \nabla - K)^2 +\frac{ \phi}{m} \r\}\psi= 2i  {\cal D} \,  \frac{\d \psi}{\d t}.
\eeq
In order to prepare the reverse derivation in which $K$ actually represents the rotational part of the velocity field of the fluid under consideration, we have given here to the potential $K$ a form in which it has the dimensionality of a velocity. In the case of an electromagnetic field, it is related to the vector potential $A$ by the relation $K=(e/mc) A$. In the particular case when ${\cal D}=\hbar/2m$, one recovers the Schr\"odinger equation of standard quantum mechanics in the presence of a vectorial field,
\beq
\l\{\frac{1}{2m} (-i \hbar \nabla -m K)^2 + \phi \r\}\psi= i \hbar \frac{\d \psi}{\d t}.
\eeq
Note that this equation may be itself founded from the scale relativistic interpretation of gauge field theories according to which the field and the charges emerge as manifestations of the fractality of space-time \cite{LN94,NCL06,LN08}. In this approach, the QED covariant derivative $-i \hbar\tilde{ \nabla}=-i \hbar \nabla -m K$ can be derived from geometric first principles, and therefore the electromagnetic Schr\"odinger equation can be established as the integral of a geodesic equation (see also \cite{JCP99}).

Let us expand the Hamiltonian. We obtain (reintroducing  for clarity indices running from 1 to 3)
\beq
-2{\cal D}^2 \Delta \psi + 2i  {\cal D} K_k \d^k \psi + i  {\cal D}(\d_k K^k) \psi + \frac{1}{2}  (K_k K^k) \psi +\frac{ \phi}{m} \psi =2 i  {\cal D}   \frac{\d \psi}{\d t}.
\eeq
We now express the wavefunction $\psi$ in terms of its modulus and of its phase,
\beq
\psi=\sqrt{P}  \times e^{i \theta}.
\eeq
Therefore we have
\beq
\d_k \psi=( \d_k \sqrt{P} + i  \sqrt{P}\, \d_k \theta) e^{i \theta}, \;\;\;\d_t \psi=( \d_t \sqrt{P} + i  \sqrt{P}\, \d_t \theta) e^{i \theta},
\eeq
\beq
\Delta \psi=\l\{ (\d_k\d^k \sqrt{P} -\sqrt{P} \, \d_k \theta\, \d^k \theta)+i(2 \, \d_k \theta\, \d^k \sqrt{P} + \sqrt{P} \, \d_k\d^k\theta)\r\}e^{i \theta}.
\eeq
The Schr\"odinger equation becomes, after simplification of the $ e^{i \theta}$ term in factor,
\begin{eqnarray}
 -2{\cal D}^2 (\d_k\d^k \sqrt{P} -\sqrt{P} \, \d_k \theta\, \d^k \theta)-2{\cal D} \sqrt{P} \, K_k \d^k \theta+ \l(\frac{1}{2}   K_k K^k+\frac{\phi}{m} \r) \sqrt{P} + 2{\cal D} \sqrt{P} \, \d_t \theta    \nonumber\\
+i \l\{ -2{\cal D}^2 \l( \sqrt{P}\,  \d_k\d^k\theta +2\,  \d_k \theta\, \d^k \sqrt{P} \r) + 2{\cal D}  K_k \,\d^k \sqrt{P}+{\cal D} (\d_k K^k) \sqrt{P}- 2{\cal D} \, \d_t \sqrt{P}  \r\} =0.
\label{2785}
\end{eqnarray}

%¡¡¡¡¡¡¡¡¡¡¡¡¡¡¡¡¡¡¡¡¡¡¡¡¡¡¡¡¡¡¡¡¡¡¡¡¡¡¡
\subsubsection{Continuity equation}

Let us first consider the imaginary part of this equation. After multiplication by $2 \sqrt{P}$ it becomes
\beq
-2  {\cal D} \, \d_t P -2  {\cal D}^2 (2 P \Delta \theta+ 2 \d_k P\,  \d^k \theta)+2  {\cal D}(K_k \d^k P+P \d_kK^k)=0.
\eeq
Without the indices, it reads
\beq
\d_t P + 2{\cal D} (P \Delta \theta + \nabla P. \nabla \theta) -K .\nabla P -P \, \nabla. K=0.
\eeq
Let us now introduce, as in the scalar field case, a potential motion velocity field 
\beq
V= 2 {\cal D} \nabla \theta.
\eeq
We obtain
\beq
\d_t P + P\,  \nabla.V -P\, \nabla. K +\nabla P.V - \nabla P. K=0.
\eeq
This leads us to define a full `velocity field' as 
\beq
v=V-K,
\eeq
in terms of which the above equation reads
\beq
\d_t P + P\,  \nabla.v +\nabla P.v=0,
\eeq
and finally becomes the continuity equation
\beq
\frac{\d P}{\d t} + {\rm div} (P \, v)=0,
\eeq
which is therefore generally valid, provided it is written in terms of the full velocity field $v=V-K$ instead of only the velocity field $V$ (which is linked to the phase of the wave function).

%¡¡¡¡¡¡¡¡¡¡¡¡¡¡¡¡¡¡¡¡¡¡¡¡¡¡¡¡¡¡¡¡¡¡
\subsubsection{Energy equation}

Let us now consider the real part of equation~(\ref{2785}).  It reads
\beq
\sqrt{P} \l[ \l(2{\cal D} \, \d_t \theta +2{\cal D}^2 \d_k \theta\, \d^k \theta-2{\cal D}^2\frac{\Delta\sqrt{P}}{\sqrt{P}}+\frac{\phi}{m} \r)-2{\cal D} \, K_k \, \d^k \theta+ \frac{1}{2}   K_k K^k \r]=0.
\eeq
We now use the equivalent notation $S= 2m {\cal D} \theta$ ($=\hbar \theta$ in the case of standard quantum mechanics), so that the wave function is now defined, like in previous sections, as
\beq
\psi=\sqrt{P} \times e^{iS/2m{\cal D}}.
\eeq
 We obtain
\beq
\sqrt{P} \l[ \d_t S +\frac{1}{2m} \d_kS\, \d^k S +\phi -2 m {\cal D}^2 \frac{\Delta\sqrt{P}}{\sqrt{P}} -K_k\, \d^k S+ \frac{1}{2}  m K_k K^k \r]=0.
\eeq
The potential part of the full velocity field now reads
\beq
 V=\frac{\nabla S}{m},
\eeq
and we get the energy equation
\beq
\frac{\d S}{\d t}+\frac{1}{2} m V^2  +\frac{1}{2} m K^2-m V.K+\phi-2 m {\cal D}^2 \frac{\Delta\sqrt{P}}{\sqrt{P}}=0.
\eeq
One recognizes, once again, the emergence of the full velocity field $v=V-K$ in this equation, where $V$ is potential while $K$ is rotationnal.
In its terms the energy equation takes the same form as in the scalar field case, namely,
\beq
-\frac{\d S}{\d t}=\frac{1}{2} m v^2+\phi-2 m {\cal D}^2 \frac{\Delta\sqrt{P}}{\sqrt{P}}.
\eeq
When the energy is conserved, $E=-\d S / \d t$. We therefore recover the same three contributions of kinetic energy $E_c=\frac{1}{2} m v^2$, exterior potential energy $\phi$, and quantum potential energy
\beq
Q=-2 m {\cal D}^2 \; \frac{\Delta\sqrt{P}}{\sqrt{P}},
\eeq
as in the previous case. The quantum potential also keeps exactly its previous form in this new (vectoriel field) situation.

Let us now take the gradient of the energy equation. One obtains
\beq
\frac{\d V}{\d t} + \frac{1}{2} \nabla (v^2 )= -\nabla \l(\frac{\phi+Q}{m}\r).
\eeq
In the potential case, $ \frac{1}{2} \nabla (v^2 )= (v.\nabla) v$. But here, in the case of rotational motion, this relation leads to the introduction of a vorticity-like quantity, $\omega= {\rm curl} \, v$, i.e.,
\beq
\omega_{\alpha k}=\d_\alpha v_k-\d_k v_\alpha=\d_k K_\alpha  -\d_\alpha K_k.
\eeq
Since $K$ represents here a vector potential,  $-\omega= {\rm curl} \, K$ therefore represents a magnetic-like field.
In tensorial notation we have
\beq
 \frac{1}{2} \d_\alpha(v^k v_k)= v^k \d_\alpha v_k= v^k \d_k v_\alpha+ v^k(\d_\alpha v_k-\d_k v_\alpha)=v^k \d_k v_\alpha+ v^k \omega_{\alpha k},
\eeq
i.e.,
\beq
 \frac{1}{2} \nabla (v^2 )=(v.\nabla) v + v \times \omega.
\eeq
Therefore, since $V=v+K$ and  $ {\rm curl} \, v=- {\rm curl} \, K$, one finally obtains the equation
\beq
\frac{\d v}{\d t} + (v.\nabla) v =-\frac{\d K}{\d t}  + v \times {\rm curl} \, K -\nabla \l(\frac{\phi+Q}{m}\r).
\eeq
One recognizes in the right-hand side of this equation the exact analog of a Lorentz force, to which is added the quantum force $ -\nabla Q/m$. The  term $-{\d K}/{\d t}$ is the analog of the magnetic contribution $-\d A/ c \, \d t$ to the electric field ${\cal E}=-\d A/ c \, \d t-\nabla \phi$, while $ v \times {\rm curl} \, K$ is the analog of the magnetic force $(e/c)\, v \times {\rm curl} \, A$ (see e.g. \cite{Landau2}). 

This equation has therefore exactly the form of the Euler equation that is expected for a fluid of velocity field $v$ coupled to a scalar potential $\phi$ and to a vectorial potential $K$, and subjected to an additional quantum potential $Q$. It agrees with the continuity equation which is also written in terms of the full velocity field $v$.

%¡¡¡¡¡¡¡¡¡¡¡¡¡¡¡¡¡¡¡¡¡¡¡¡¡¡¡¡¡¡¡¡¡¡¡¡¡¡¡
\subsubsection{From Ginzburg-Landau equation to fluid equations with magnetic field and quantum potential}
\label{GL}

Such an approach can be applied to the transformation of the Ginzburg-Landau equation of superconductivity into the classical equations for a fluid subjected to a magnetic field and to a quantum-like potential.

Let us start indeed from the Ginzburg-Landau equation of superconductivity \cite{Landau9} generalized to a coefficient $\cal D$ which may be different from $\hbar/2$,
\beq
\l( {\cal D} \nabla -i \frac{K}{2}\r)^2 \psi+ \alpha \, \psi  - \beta |\psi|^2 \, \psi=0,
\eeq
where $A=(mc/e)K$ is the magnetic vector potential.

From the previous decomposition, it is equivalent to the classical continuity and Euler equations of a fluid subjected both to a Lorentz force and to a quantum potential $Q$, namely (for $m=1$)
\beq
\frac{\d P}{\d t} + {\rm div} (P \, v)=0,
\eeq
\beq
\frac{\d v}{\d t} + v.\nabla v =-\frac{\d K}{\d t}  + v \times {\rm curl} \, K -\nabla Q,
\eeq
where $P=|\psi|^2$ and
\beq
Q=-2 {\cal D}^2 \; \frac{\Delta\sqrt{P}}{\sqrt{P}}.
\eeq
The reversibility of the transformation (see next section~\ref{vorticity}) means that, if one applies to a classical charged fluid a classical force having exactly the form of the `quantum potential' $Q$ (with a coefficient $\cal D$ no longer limited to the microscopic value $\hbar/2$), such a fluid would be described by the Ginzburg-Landau equation and it would therefore acquire some of the properties of a superconductor.

%¡¡¡¡¡¡¡¡¡¡¡¡¡¡¡¡¡¡¡¡¡¡¡¡¡¡¡¡¡¡¡¡¡¡¡¡¡¡¡¡¡¡¡¡¡¡¡¡¡¡¡¡¡¡¡¡¡¡¡
\subsubsection{Euler equation when $ v \times {\rm curl} \, v$ vanishes}

A more simple form of Euler equation may be recovered in rather general situations, as we shall now see.

When $ v \times {\rm curl} \, v=0$, this means that $v$ and ${\rm curl} \, v$ are parallel, i.e., ${\rm curl} \, v= \lambda \, v$ (Beltrami stream). In this case the Schr\"odinger in vectorial field equation takes the form of a standard Euler and continuity system of equations for a fluid subjected to a quantum-type potential $Q=-2 m {\cal D}^2 \; {\Delta\sqrt{P}}/{\sqrt{P}}$ and to a force $F_K=-{\d K}/{\d t}$, namely,
\beq
\frac{\d v}{\d t} + (v.\nabla) v = F_K   -\nabla \l(\frac{\phi+Q}{m}\r),
\eeq
\beq
\frac{\d P}{\d t} + {\rm div} (P \, v)=0.
\eeq

%¡¡¡¡¡¡¡¡¡¡¡¡¡¡¡¡¡¡¡¡¡¡¡¡¡¡¡¡¡¡¡¡¡¡¡¡¡¡¡¡¡¡¡¡¡¡¡¡¡¡¡¡¡¡¡¡¡¡¡
\subsubsection{Euler equation when $ v \times {\rm curl} \, v$ is a gradient}

When $ v \times {\rm curl} \, v=\nabla \xi_f/m$,which corresponds to  ${\rm curl} (v \times {\rm curl} \, v)=0$, $ \xi_f$ plays the role of an additional scalar potential, and the Schr\"odinger equation in vectorial field may also be given the form of a standard Euler and continuity system of equations for a fluid subjected to a quantum-type force $F_Q=- \nabla Q$, with $Q=-2 m {\cal D}^2 \; {\Delta\sqrt{P}}/{\sqrt{P}}$, to a force $F_K=-{\d K}/{\d t}$, and to a total force $F=-\nabla( \xi_f+\phi)/m$, namely,
\beq
\frac{\d v}{\d t} + (v.\nabla) v = F_K   -\nabla \l(\frac{\phi+ \xi_f+Q}{m} \r),
\eeq
\beq
\frac{\d P}{\d t} + {\rm div} (P \, v)=0.
\eeq

%1.6**************************************************************************
\subsection{Inverse problem: from Euler equation with vorticity to Schr\"odinger equation in vectorial field}
\label{vorticity}

The previous calculations are reversible in the case when $ v \times {\rm curl} \, v$ is a gradient, and therefore they allow us to achieve a new result. Namely, the equations of the motion of a fluid including a rotational component subjected to a quantum-type potential can also be integrated in terms of a (possibly non-linear) Schr\"odinger equation, the rotational part of the motion appearing in it under the same form as an external vectorial field. 

Consider a classical non-viscous fluid subjected to a scalar potential $\phi$ and described by its velocity field $v(x,y,z,t)$ and its density $\varrho(x,y,z,t)$. These physical quantities are solutions of the Euler and continuity equations,
\beq
\l(\frac{\d}{\d t} + v.\nabla\r)v= -{\nabla}\phi  -\frac{\nabla p} {\varrho}.
\eeq
\beq
\frac{\d \varrho}{\d t} + {\rm div} (\varrho \, v)=0.
\eeq
In the case of an isoentropic fluid, and more generally in all cases when there exists a univocal link between the pressure $p$ and the density $\varrho$, ${\nabla p}/ {\varrho}$ becomes a gradient \cite{Landau6}, namely ${\nabla p}/ {\varrho}=\nabla w$, where $w$ is the enthalpy by mass unit in the isentropic case ($s=$ cst). In this case we set
\beq
 {\nabla}\phi  +\frac{\nabla p} {\varrho}= \nabla(\phi +w)=\nabla \Phi,
\eeq
and the Euler equation becomes
\beq
\l(\frac{\d}{\d t} + v.\nabla\r)v= -{\nabla}\Phi.
\eeq

Let us now assume that the classical fluid is sujected to an additional force
\beq
F_Q=-\nabla Q=2 {\cal D}^2 {\nabla} \l( \frac{\Delta \sqrt{\varrho}}{\sqrt{\varrho}}\r),
\eeq
so that the Euler and continuity equations read
\beq
\label{1230}
\l(\frac{\d}{\d t} + v.\nabla\r)v= -{\nabla}\l(w+\phi+Q \r),
\eeq
\beq
\label{1231}
\frac{\d \varrho}{\d t} + {\rm div} (\varrho \, v)=0.
\eeq
The Euler equation can be written under the form
\beq
\frac{\d v}{\d t} + \frac{1}{2} \nabla( v^2) - v \times {\rm curl} \, v =  -\nabla (w+\phi+Q),
\eeq
In general the velocity field $v$ is not potential. Let us decompose it in terms of a potential (irrotational) contribution $V$ and a rotational one $K$. Namely, we set
\beq
v=V-K, \;\;\; V= \nabla S.
\eeq
Then we build a ``wavefunction" by combining this function $S$ and the density $\varrho$ in terms of a complex function:
\beq
\psi=\sqrt{\varrho} \times e^{i S/2{\cal D}}.
\eeq
Therefore $\d v/\d t=\d \nabla S/\d t-\d K /\d t=\nabla(\d S/\d t)-\d K /\d t$, so that the Euler equation now reads
\beq
\frac{\d K}{\d t} - v \times {\rm curl} \, K =  \nabla \l(\frac{\d S}{\d t}+ \frac{1}{2} v^2+w+\phi+Q \r).
\eeq
In the general case, the left-hand side of this equation is not a gradient, so that it cannot be integrated under the form of a (generalized) Schr\"odinger equation. 

However, there is a large class of configurations when this is possible. Indeed, in the case when
\beq
\frac{\d K}{\d t}  - v \times {\rm curl} \, K=-\nabla \chi,
\eeq
the Euler equation can be integrated in terms of an energy equation that reads
\beq
\frac{\d S}{\d t}+ \frac{1}{2} v^2+w+\phi+\chi+Q=0,
\eeq
up to a constant which can be set to zero by a gauge transformation of the phase $S$.
We have therefore recovered  the conditions (energy equation and continuity equation) which lead to the construction of a nonlinear (NL) Schr\"odinger-type equation. 

Therefore the whole calculation of the previous Section~\ref{sec2.5} can be reversed in this case (with $m=1$ and $P\propto \varrho$), so that  we can integrate the Euler and continuity system in terms of a non-linear ``magnetic" Schr\"odinger-type equation that writes
\beq
\l( {\cal D} \nabla -i \frac{K}{2}\r)^2 \psi+  i  {\cal D} \,  \frac{\d \psi}{\d t}=\l( \frac{w+\phi+\chi}{2} \r)\,\psi,
\label{ooo}
\eeq
where $\psi = \sqrt{\varrho}\times \exp(i S/ 2 {\cal D})$. In general the pressure, and therefore the enthalpy $w$ is a function of the density $\rho=|\psi|^2$, which contributes to the non-linearity of this equation. In the absence of vorticity, it is similar to the kind of NL Schr\"odinger equation encountered in the study of superfluids and Bose-Einstein condensates  (see e.g. \cite{Fetter1965, Nore}). 

Equation~(\ref{ooo}) can also be given an expanded form,
\beq
 {\cal D}^2 \Delta \psi +  i  {\cal D} \,  \frac{\d \psi}{\d t}=\l\{ \frac{\phi +w +\chi}{2} + \frac{K^2}{4}+i \,\frac{\cal D} {2}\, \nabla . K  +i {\cal D} K . \nabla \r \}\,\psi,
\eeq
where the term between brackets in the right-hand side may be interpreted, when $K . \nabla \psi$ is negligible, as a generalized potential energy.

The vorticity  $\omega=-{\rm curl}\, K$ plays a role similar to a vectorial field.  Its evolution equation, which can be obtained by taking the curl of the Euler equation, namely,
\beq
\frac{\d \omega}{\d t} ={\rm curl} (v \times \omega),
\eeq
becomes identically verified in the considered case when ${\d K}/{\d t}  - v \times {\rm curl} \, K$ is a gradient. But it is no longer here an independent field externally applied, so that the velocity field $K$ in the Schr\"odinger equation is in this case one of the unknowns to be solved. Nevertheless, despite this situation, this result may remain physically meaningful and useful, since the Schr\"odinger equation and its solutions has general properties which are valid whatever the applied fields.

Let us finally consider the stationary version of Eq.~(\ref{ooo}) in the general case when the pressure terms reads $w=p/\rho \propto \rho$ (see Sec.~\ref{pressure}). We obtain
\beq
\l( {\cal D} \nabla -i \frac{K}{2}\r)^2 \psi+ \alpha \, \psi  - \beta |\psi|^2 \, \psi=0,
\eeq
with $\alpha=(E-\phi-\chi)/{2}$. This equation has exactly the form of the Ginzburg-Landau equation of superconductivity \cite{Landau9}, generalized to a coefficient $2\cal D$ which may be different from $\hbar$. This result may be applied to the two cases initially considered at the beginning of Sec.~\ref{sec2.5}, namely:

\noindent(i) A classical charged fluid subjected to an electromagnetic field and to a classical potential taking the form of the quantum potential $Q$. As already remarked in Sec.~\ref{GL}, the equations of motion of such a fluid (continuity equation and Euler equation with a Lorentz force) may be combined in terms of a single complex equation which takes the form of the Ginzburg-Landau equation of superconductivity. One may therefore hope such a fluid to acquire some of the properties of a quantum fluid.

\noindent(ii)  The potential $K$ does not represent here an external magnetic field, but a rotational part of the velocity field. This means that a non-linear Schr\"odinger form can also be given to some fluids showing vorticity and subjected to an external potential having a quantum-like form.

\subsection{From Navier-Stokes to non-linear Schr\"odinger equation}
%*************************************************************************************************
Let us finally consider the general case of Navier-Stokes equations including a viscosity term.
The fluid mechanics equations including a quantum-type potential read in this case
\beq
\l(\frac{\d}{\d t} + v.\nabla\r)v= \nu \Delta v  -\frac{\nabla p}{\varrho}-\nabla(\phi+Q),
\eeq
\beq
\frac{\d \varrho}{\d t} + {\rm div} (\varrho \, v)=0,
\eeq
where the quantum-type potential energy is still given by
\beq
Q=-2{\cal D}^2 \frac{\Delta \sqrt{\varrho}}{\sqrt{\varrho}}.
\eeq
We set as in the previous section
\beq
v=V-K, \;\;\; V= \nabla S,\;\;\;  \psi=\sqrt{\varrho} \times e^{i S/2{\cal D}},
\eeq
i.e., $V$ is the potential part of the full velocity field $v$. Therefore the viscosity term reads
\beq
\nu \Delta v=\nu \Delta(\nabla S-K)= \nu \nabla(\Delta S) - \nu \Delta K,
\eeq
and, assuming once again ${\nabla p}/{\varrho}=\nabla w$, the Navier-Stokes equation now takes the form
\beq
-\frac{\d K}{\d t}+\nu \Delta K  - v \times {\rm curl} \, v =  -\nabla \l(\frac{\d S}{\d t}-\nu \Delta S+ \frac{1}{2} v^2+w+\phi+Q \r).
\eeq
This equation is not generally integrable. However, it nevertheless becomes integrable for a large class of flows, namely, those for which its left-hand side is a gradient,
 \beq
 -\frac{\d K}{\d t}+\nu \Delta K  - v \times {\rm curl} \, v=\nabla \chi.
 \eeq
 In this case one obtains an energy and a continuity equation that read
 \beq
 \frac{\d S}{\d t}-\nu \Delta S+ \frac{1}{2} v^2+w+\phi+\chi+Q=0,\;\;\;\; \frac{\d \varrho}{\d t} + {\rm div} (\varrho \, v)=0,
 \eeq
and which can be combined into the form of a non-linear Schr\"odinger equation of the magnetic type,
\beq
\l( {\cal D} \nabla -i \frac{K}{2}\r)^2 \psi+  i  {\cal D} \,  \frac{\d \psi}{\d t}= \frac{1}{2}  \l(w+\phi+\chi-\nu \Delta S \r)\,\psi.
\eeq
The viscosity therefore leads to add a new nonlinear term in this NL Schr\"odinger equation, that depends on the phase $S/2{\cal D}$ of the wave function. When the fluid motion is irrotational, the integration under the form of a NL Schr\"odinger equation of the continuity and Navier-Stokes equations including a quantum potential is always possible. 

%*******************************************************************
\section{Schr\"odinger equation for the rotational motion of a solid}
%*******************************************************************

%*************************
\subsection{Introduction}
%************************

In the previous sections, a Schr\"odinger form has been obtained for the equations of motion and of continuity of a fluid subjected to a quantum potential.  However, the method we used may be applied not only to a fluid but also to a mechanical system. Indeed, we have shown \cite{LN97} that the scale relativity approach can be applied to the rotational motion of a solid, leading once again to a Schr\"odinger-type equation. Here we give an improved demonstration of this Sch\"odinger equation, then, as in previous sections, decompose it in terms of its real and imaginary parts, and then obtain a new generalized form of the quantum potential. Reversely, the addition of such a new quantum potential in the energy equation yields, in combination with the continuity  equation, a Schr\"odinger equation.

%******************************************************
\subsection{Equation of rotational solid motion in scale relativity}
%******************************************************
Let us briefly recall the results of  \cite{LN97,LN98}, in which a Schr\"odinger form was obtained for the equation of the rotational motion of a solid subjected to the three basic effects of a fractal and nondifferentiable space (namely, infinity of trajectories, fractal dimension 2 and reflection symmetry breaking of the time differential element).

The role of the variables $(x, v, t)$ of translational motion is now played respectively by $(\varphi, \Omega, t)$, where $\varphi$ stands for three rotational position angles (for example, Euler angles)  and $\Omega$ for the angular velocity. We choose a contravariant notation $\varphi^k$ for the angles, where the indices run from 1 to 3, and we adopt Einstein's convention for summation on upper and lower indices. The Euler-Lagrange equations for rotational motion classically write \cite{Landau1}
\beq
\frac{d}{dt} \, \frac{\d L}{\d \Omega^k}= \frac{\d L} {\d \varphi^k},
\eeq
in terms of a Lagrange function $L=(1/2) I_{ik}\Omega^i \Omega^k - \Phi$, where $I_{ik}$ is the tensor of inertia of the solid body and $\Phi$ its potential energy in an exterior field. Therefore the angular momentum of the system is 
\beq
M_i= \frac{\d L}{\d \Omega^i}=I_{ik}\Omega^k.
\eeq
The torque  is given by
\beq
K_i=\frac{\d L} { \d  \varphi^i}=-\frac{\d \Phi}{\d \varphi^i},
\eeq
and the motion equations finally take the Newtonian form
\beq
\frac{dM_i}{dt}= K_i.
\eeq

Let us now consider the generalized description of such a system in the scale relativity framework. Following the same road as for position coordinates, in the generalized situation when space-time is fractal, the angle differentials $d\varphi= dx_{\varphi}+d \xi_{\varphi}$ can be decomposed in terms of two contributions, a classical (differentiable) part  $dx_{\varphi}$ and a fractal fluctuation $d \xi_{\varphi}$ which is such that $\langle d \xi_{\varphi} \rangle=0$ and
\beq
\langle d \xi_{\varphi}^j \, d \xi_{\varphi}^k \rangle = 2 {\cal D}^{jk} dt.
\eeq
where ${\cal D}^{jk}$ is now a tensor which generalizes the scalar parameter $\cal D$ of the translational case. As we shall see in the following, this tensor is, up to a multiplicative constant, similar to a metric tensor. 

The breaking of reflexion invariance ($dt \leftrightarrow -dt$) on the time differential elements, which is a consequence of the nondifferentiability, yields
a two-valuedness of the angular velocity  \cite{LN93,LN96,CN04}. This leads to introducing a complex angular velocity $\widetilde{\Omega}$, then a complex Lagrange function $\widetilde{L}(\varphi, \widetilde{\Omega}, t)$. The two effects of nondifferentiability and fractality of space can finally be combined in terms of a rotational quantum-covariant derivative \cite{LN97},
\begin{equation}
\label{F}
\frac{\dfr}{dt}  =  \frac{\partial}{\partial t} + \widetilde{\Omega}^{k} \, \partial_{k}
- i \,{\cal D}^{jk} \partial_{ j} \partial_{ k},
\end{equation}
where $\d_k=\d/\d \varphi^k$.
Using this quantum-covariant derivative, we may generalize to fractal motion the equation of rotational motion while keeping its classical form,
\begin{equation}
\label{G}
I_{ jk} \; \frac{\dfr \:{\widetilde{\Omega}^{k}}}{dt} = -\partial_{ j} \Phi,
\end{equation}
where $I_{ jk}$ is the tensor of inertia of the solid and $\Phi$ an externally added potential.

We then introduce a complex function (which will subsequently be identified with a wave function) as another expression for the complex action $\widetilde{S} = \int{\widetilde{L} \, dt}$,
\begin{equation} 
\label{G1}
\psi = e^{i \widetilde{S}/S_0},
\end{equation}
where $S_0$ is a real constant introduced for dimensional reasons. Now the complex angular momentum is, like in classical solid mechanics,  linked to the complex action by the standard relation $\widetilde{M}_{k}=\d \widetilde{S}/ \d \varphi^k$, so that one obtains
\beq
\widetilde{M}_{k}=I_{\alpha k} \widetilde{\Omega}^k=-i S_0 \, \d_{\alpha} \ln \psi.
\eeq
Let us therefore introduce the inverse of the tensor of inertia, $[I]^{-1}=I^{\alpha k}$, such that
\beq
I_{\alpha k} I^{k \beta}= \delta_{\alpha}^{\beta}.
\eeq
This allows us to express the complex velocity field in terms of the wave function,
\beq
\widetilde{\Omega}^{k}=-i S_0 \, I^{k\alpha }\, \d_{\alpha} \ln \psi.
\eeq
We can now replace the velocity field by this expression in the covariant derivative and in the rotational motion equations. We obtain
\beq
-i S_0 \l( \frac{\d}{\d t} -i S_0  \, I^{k\beta } \d_{\beta} \ln \psi \, \d_k -i\, {\cal D}^{jk} \partial_{ j} \partial_{ k} \r) \d_{\alpha} \ln \psi= K_{\alpha},
\eeq
which can be written as
\beq
-i S_0 \l( \d_{\alpha} \frac{\d}{\d t}\ln \psi-i \{ S_0  \:  \d_{\beta} \ln \psi \:  I^{k\beta}  \d_k \, \d_{\alpha} \ln \psi+{\cal D}^{jk} \partial_{ j} \partial_{ k} \, \d_{\alpha} \ln \psi \}\r)= K_{\alpha}.
\eeq
We have reversed in the second equation the places of $I^{k\beta }$ and $\d_{\beta}$: this is possible since $I^{k\beta }$ is assumed to be constant. This reversal allows one to make appear the operator $I^{k\beta}  \d_k$. Provided the tensor of inertia plays the role of a metric tensor, we have $I^{k\beta}  \d_k= \d^\beta$, and we recognize in the expression under brackets $\{ \}$ a tensorial generalization of the expression which was encountered in the translational motion case, namely,
\beq
 S_0 \,m^{-1}  \,( \d_{\beta} \ln \psi \, \d^{\beta}) \d_{\alpha} \ln \psi+{\cal D} \partial_{k} \partial^{ k}  \d_{\alpha} \ln \psi.
\eeq
Indeed, as recalled at the beginning of this paper, the relation
\beq
S_0 \,m^{-1}= 2{\cal D},
\eeq
that is nothing but a generalized Compton relation, allows one to transform this expression into a remarkable identity which leads to the integration of the motion equation in terms of a Schr\"odinger equation. 

Now we are able to generalize this result to the rotational motion case, despite the complication brought by the fact that the mass is replaced by the inertia tensor. Indeed, the inverse of the mass is replaced by the inverse tensor $[I]^{-1}$, and  we can identify the fractal fluctuation tensor with this metric tensor up to a constant, namely
\beq
{\cal D}^{\alpha \beta }=\frac{S_0}{2} \, I^{\alpha \beta }  ,
\eeq
i.e., in matrix form, $S_0 [I]^{-1}= 2 [{\cal D}]$. (Note the correction to \cite{LN97,LN98} where an inverse relation between these quantities was erroneously given; the Schr\"odinger equation obtained in these papers nevertheless remains correct). This is a new tensorial generalization of the Compton relation. Moreover this means that it is the inertia tensor itself which serves as a metric tensor and can be used to raise and lower the indices, e.g., $I^{jk} \partial_{ j} \partial_{ k}=\d^k \d_k$, while ${\cal D}^{jk}$ does the same but up to a constant, namely, ${\cal D}^{jk}\partial_{ j} \partial_{ k}=(S_0/2)\d^k \d_k$.

The existence of a similarity between the rotational diffusion term $\widehat{M}_j D^{jk} \widehat{M}_k$, where $\widehat{M}$ denotes angular momentum operators, and the corresponding quantum mechanical Hamiltonian $\widehat{M}_j I^{jk} \widehat{M}_k /2$ of a rigid body has already been remarked by Dale Favro \cite{DaleFavro1960} in his theory of rotational Brownian motion. Here we  directly identify ${\cal D}^{jk}$ to $(S_0/2)I^{jk}$ (but ${\cal D}^{jk}$, despite its stochastic definition, should not be confused with a standard diffusion coefficient).

The equation of motion now takes the form
\beq
-iS_0 \l( \d_{\alpha} \frac{\d}{\d t}\ln \psi-i \frac{S_0}{2} \{ I^{k \beta }  \d_{\beta} \ln \psi \, \d_k  \d_{\alpha} \ln \psi+{I}^{jk} \partial_{ j} \partial_{ k}  \d_{\alpha} \ln \psi \}\r)= K_{\alpha},
\eeq
and, using the tensorial notation ${I}^{jk} \partial_{ j}= \partial^{k}$, it can now be written as
\beq
-iS_0 \l( \d_{\alpha} \frac{\d}{\d t}\ln \psi-i\frac{S_0}{2} \{ 2  \d^{k} \ln \psi \, \d_k  \d_{\alpha} \ln \psi+\partial^{k} \partial_{ k}  \d_{\alpha} \ln \psi \}\r)= K_{\alpha}.
\eeq
We recognize in this expression the remarkable identity Eq.~(\ref{idgen}) of Sec.~\ref{ident} so that we can simplify it under the form
\beq
-iS_0 \l( \d_{\alpha} \frac{\d}{\d t}\ln \psi-i\frac{S_0}{2} \d_\alpha \frac{\partial_{k} \partial^{k} \psi}{\psi}\r)=-\d_\alpha \Phi,
\eeq
and finally write it globally as a gradient,
\beq
\d_\alpha S_0 \l\{ \frac{({S_0}/{2})\partial_{k} \partial^{k} \psi+ i \d \psi/ \d t}{\psi} \r\}= \d_\alpha \Phi.
\eeq
This equation can therefore be  integrated in the general case under the form of a new generalized Schr\"odinger equation that reads \cite{LN97,LN98}
\begin{equation}
S_0  \l({\cal D}^{ jk} \partial_{ j} \partial_{ k} \psi + i\,   \frac{\partial}{\partial t} \psi \r) =  \Phi \, \psi.
\end{equation}
In terms of the inverse tensor of inertia this rotational Schr\"odinger equation reads
\beq
\frac{1}{2}S_0^2 \, I^{ jk} \partial_{ j} \partial_{ k} \psi + i\, S_0\,  \frac{\partial}{\partial t} \psi =  \Phi \, \psi.
\eeq
Since the tensor of inertia plays the role of a metric tensor, in particular for the lowering and raising of indices, it can finally be written as
\beq
\frac{1}{2}S_0^2 \:  \partial^{ k} \partial_{ k} \psi + i\, S_0\,  \frac{\partial}{\partial t} \psi =  \Phi \, \psi,
 \label{H}
\eeq
which keeps the form of the scalar case \cite{LN93}, while generalizing it.

The standard quantum case is recovered by identifying $S_0$ with $\hbar$, but, once again, all the mathematical structure of the equation (and therefore of its solutions) is preserved with a constant that can have any value, including a macroscopic one.

We may now conclude by returning to the fractal angular fluctuations, that writes in terms of the inverse inertia tensor
\beq
\langle d \xi_{\varphi}^j \, d \xi_{\varphi}^k \rangle = S_0 {I}^{jk} dt.
\eeq
We therefore gain a complete justification of the identification of the tensor of inertia with a metric tensor, since, owing to the fact that $I^{kj}I_{jk}=\delta_k^k=3$, we obtain the invariant metric relation
\beq
dt =\frac{ {I}_{jk}}{3 S_0} \langle d \xi_{\varphi}^j \, d \xi_{\varphi}^k \rangle,
\eeq
where $S_0=\hbar$ in the standard quantum case, and where $dt$ (which appears instead of its square $dt^2$ as an expression of the fractal dimension 2) is indeed the fundamental invariant here since all this study is done in the framework of Galilean motion relativity.

\subsection{Fluid representation and newly generalized quantum potential}
%*******************************************************************************

Let us now give this Schr\"odinger equation a fluid mechanical form. This can be easily done by following the same steps as in Sec.~\ref{sec3.1}, but now using tensorial derivative operators. We set
\beq
\psi= \sqrt{P} \times e^{i S/S_0},
\eeq
and we replace $\psi$ by this expression in Eq.~(\ref{H}). The imaginary part of this equation reads
\beq
\frac{1}{2}  \sqrt{P}\, \d_k \d^k {S}+  \, \d_k {S}\, \d^k \sqrt{P} + \frac{\d \sqrt{P}}{\d t}=0.
\eeq
i.e., after simplification,
\beq
\frac{\d P}{\d t} + \d_k(P\d^k S)=0.
\eeq
Since $S$ is the real part of the complex action, it is linked to the angular momentum $M^k$ (which is itself the real part of the complex angular momentum) and to the real angular velocity $\Omega^j$  by the relations
\beq
M_\alpha=I_{\alpha j} \Omega^j=\d_\alpha S.
\eeq
Now, since $I^{k \alpha} I_{\alpha j}=\delta_k^j$, we find
\beq
\Omega^k=I^{k \alpha} I_{\alpha j} \Omega^j=I^{k \alpha}M_\alpha.
\eeq
Therefore
\beq
\Omega^k=I^{k \alpha} \d_\alpha S= \d^k S.
\eeq
Finally, we find that the imaginary part of the rotational Schr\"odinger equation amounts, once again, to a continuity equation in the general case
\beq
\frac{\d P}{\d t} +\d_k(P \Omega^k)=0.
\eeq
Note the correction to Refs.~\cite{LN97,LN98}, in which we concluded that this was the case only in some particular reference systems. This means that the probability interpretation of $P=|\psi|$ is also generally ensured  \cite{CN04,LN06}. 

The real part of Eq.~(\ref{H}) takes the form
\beq
\frac{\d S}{\d t}+ \Phi - \frac{1}{2}S_0^2 \frac{ \partial_{ k} \partial^{ k} \sqrt{P}}{\sqrt{P}} + \frac{1}{2}\partial_{k} S \, \partial^{ k}S=0.
\eeq
Let us first consider the last term of this expression. It reads
\beq
\frac{1}{2} \d_k S \, \d^k S=\frac{1}{2}M_k \, \Omega^k =\frac{1}{2}I_{jk} \, \Omega^j \Omega^k=T_{\rm rot},
\eeq 
which is the classical expression of the rotational kinetic energy. In the conservative case, $E=-\d S / \d t$ is the total energy, so that we recover the standard energy equation, 
\beq
E= \Phi +Q+ T_{\rm rot},
\eeq
but which now includes an additional potential energy that reads
\beq
Q= - S_0\, \frac{{\cal D}^{ jk} \partial_{ j} \partial_{ k} \sqrt{P}}{\sqrt{P}}= -\frac{1}{2} S_0^2 \,  \frac{{ I}^{ jk} \partial_{ j} \partial_{ k} \sqrt{P}}{\sqrt{P}}.
\label{rotQ}
\eeq
This is a new generalization of the quantum potential in the rotational case. 

An Euler-like equation including this quantum potential is simply obtained by taking the gradient of this equation, namely,
\beq
\d_\alpha \l( \frac{\d S}{\d t} \r) + \d_\alpha \l( \frac{1}{2} \Omega^k M_k   \r)= -\d_\alpha( \Phi+Q).
\eeq
Now one has $ \d_\alpha (\d S / \d t)= \d(\d_\alpha S)/\d t= \d M_\alpha/\d t$, and since $M_\alpha= \d_\alpha S$ is a gradient, $\d_\alpha T_{\rm rot}=\Omega^k \d_\alpha M_k=\Omega^k \d_k M_\alpha$, so that we finally obtain
\beq
 \l( \frac{\d}{\d t}+ \Omega^k \d_k \r)M_\alpha= -\d_\alpha( \Phi+Q),
\eeq
which is indeed the expected generalization in terms of an Euler equation of the equation of dynamics $dM/dt=K$ in the case when the rotational velocity becomes a velocity field, namely, $\Omega=\Omega[\varphi(t),t]$.

\subsection{From Euler to Schr\"odinger equation}
%*****************************************************
Reversely, one may now consider a rotating body or an ensemble of rotating bodies which are subjected to a fluctuating motion of rotation, such that the rotational velocity can be replaced, at least as an approximation, by a rotational velocity field $\Omega=\Omega[\varphi(t),t]$. 

Assume, moreover, that each body is subjected, in addition to the torque $ -\d_\alpha \Phi$ of an exterior field, to a quantum-like torque $K_Q=-\nabla_\varphi Q$, where the quantum potential $Q$ is given by Eq.~(\ref{rotQ}). 

Such a system would be described by an Euler equation and a continuity equation,
\beq
 \l( \frac{\d}{\d t}+ \Omega^k \d_k \r)M_\alpha= -\d_\alpha \l( \Phi -\frac{1}{2} S_0^2 \,  \frac{{ I}^{ jk} \partial_{ j} \partial_{ k} \sqrt{P}}{\sqrt{P}}\r),
\eeq
\beq
\frac{\d P}{\d t} +\d_k(P \Omega^k)=0,
\eeq
which, after introducing the wave function $\psi=\sqrt{P} \times e^{i S/S_0}$, can be recombined to yield a generalized Schr\"odinger equation that reads
\beq
\frac{1}{2}S_0^2 \, I^{ jk} \partial_{ j} \partial_{ k} \psi + i\, S_0\,  \frac{\partial}{\partial t} \psi =  \Phi \, \psi,
\eeq
so that it would be expected to show some kind of quantum-type properties. Indeed, in the particular case $S_0=\hbar$, we recover the standard Schr\"odinger equation of the quantum mechanical description of a rigid body which is used, e.g., for determining the rotational levels of molecules taken as a whole. In the macroscopic case, such an equation has been applied with positive results to the study of the probability distribution of the inclination and obliquity of chaotic astronomical bodies \cite{DRN03,LN98}.

%*****************************************************************
\section{Diffusion potential opposite to the quantum potential}
\label{s:diff}
%****************************************************************

The question of the relation of the quantum theory with diffusion processes has been posed for long. Our aim in this section is not to study in detail this question, but to enlighten it by a new result, according to which a diffusion process may also be written in terms of a a Euler equation including an additional potential energy, which is exactly the {\it opposite} of a quantum potential. Such a result leads to characterize the quantum behavior, which is in many cases self-organizing and stabilizing (as demonstrated by the existence of stationary solutions describing stable structures such as atoms and molecules), as an opposite of the diffusion behavior, which is instead linked to  entropy increase and, most of the time, desorganization.

Let us consider a classical diffusion process. Such a process is described by the Fokker-Planck equation:
\beq
\frac{\d P}{\d t}+ {\rm div}(P v)=D \Delta P,
\label{FP}
\eeq
where $D$ is the diffusion coefficient. When there is no global motion of the diffusing fluid or particles ($v=0$), the Fokker-Planck equation is reduced to the usual diffusion equation for the probability $P$:
\beq
\frac{\d P}{\d t}=D \Delta P.
\eeq
This well-known equation holds for the propagation of heat (in this case $P$ is replaced by the temperature), for the diffusion of a fluid in a mixing of fluids (in this case $P$ is replaced by the concentration of the diffusing fluid), and for the Brownian motion of particles diffusing in a fluid.

Conversely, when the diffusion coefficient vanishes, the Fokker-Planck equation is reduced to the continuity equation, 
\beq
\frac{\d P}{\d t}+ {\rm div}(P v)=0.
\eeq

\subsection{Continuity equation}
%*********************************
Let us now make the change of variable:
\beq
V=v-D \nabla \ln P.
\eeq
Let us first prove that, in the general case $v\neq0$, the new velocity field $V(x,y,z,t)$ is now solution of the standard continuity equation.  Indeed, we obtain, by replacing $V$ by its expression,
\beq
\frac{\d P}{\d t} + {\rm div} (P V)=\frac{\d P}{\d t}+ {\rm div} (P v) -D \,  {\rm div} (P\nabla \ln P)=\frac{\d P}{\d t}+ {\rm div} (P v) -D \Delta P.
\eeq
Finally using the Fokker-Planck equation, we find
\beq
\frac{\d P}{\d t} + {\rm div} (P V)=0.
\eeq
\subsection{Euler equation for $v=0$}
%*********************************
Let us now establish the form of  the Euler equation for the new velocity field $V$.
Let us calculate its total time derivative, at first in the simplified case $v=0$:
\beq 
\frac{d V}{dt}= \l( \frac{\d}{\d t} + V. \nabla\r)V=- D\,\frac{\d}{\d t} \nabla \ln P + D^2 (\nabla \ln P. \nabla) \nabla \ln P.
\label{7410}
\eeq
Now, since ${\d}\nabla \ln P /{\d t}=\nabla {\d} \ln P /{\d t}=\nabla (P^{-1}{\d} P /{\d t})$, we can make use of the diffusion equation so that we obtain:
\beq
 \l( \frac{\d}{\d t} + V. \nabla\r)V=- D^2 \l(   \nabla \frac{\Delta P}{P}-(\nabla \ln P. \nabla) \nabla \ln P  \r).
 \label{2796}
\eeq
 We shall now use again the remarkable identity Eq.~(\ref{remid}),
 \beq
\frac{1}{\alpha} \; \nabla\left(\frac{\Delta  R^{\alpha}}{R ^{\alpha}}\right)=   \Delta (\nabla \ln R)+2\alpha (\nabla \ln R . \nabla )
(\nabla \ln R ).
 \label{remid1}
\eeq
 By using it for $R=P$ and $\alpha=1$, we can replace $  \nabla ({\Delta P}/{P})$ by $ \Delta (\nabla \ln P)+2 (\nabla \ln P . \nabla )\nabla \ln P$, so that Eq.~(\ref{2796}) becomes
 \beq
  \l( \frac{\d}{\d t} + V. \nabla\r)V=- D^2 \l[   \Delta (\nabla \ln P)+(\nabla \ln P. \nabla) \nabla \ln P  \r].
 \eeq
The right-hand side of this equation comes again under the identity (\ref{remid}), but now for $\alpha=1/2$. Therefore we finally obtain the following form for the Euler equation of the velocity field $V$:
\beq
  \l( \frac{\d}{\d t} + V. \nabla\r)V=- 2 D^2 \nabla \l( \frac{\Delta \sqrt{P}}{\sqrt{P}}\r).
\eeq
This result is remarkable for several reasons:

(i) It gives an equivalence between a standard fluid subjected to a force field and a diffusion process.

(ii) The above  ``diffusion force" derives from a potential $\phi_{\rm diff}=2D^2 {\Delta \sqrt{P}}/{\sqrt{P}}$. This expression introduces a square root of probability in the description of a classical diffusion process.

(iii) This ``diffusion potential" is exactly the opposite of the quantum potential $Q/m=-2{\cal D}^2 {\Delta \sqrt{P}}/{\sqrt{P}}$.

The relation between quantum-type processes and diffusion processes is now enlightened in a new way: they appear as exactly opposite, so that quantum-type behavior can be considered as an  `anti-diffusion'  process in this context. 

The change of sign of the potential has therefore dramatic consequences, since in one case it yields a classical diffusion equation which is known to lead to desorganization, irreversibility and spreading in $\sqrt{t}$ while in the other it yields a Schr\"odinger equation that allows stationary solutions and leads to structuring and self-organization.

\subsection{Euler equation for $v\neq0$}
%******************************************
Let us now consider the general situation of a non-vanishing global velocity field $v$. In order to do this calculation we now introduce the indices in an explicit way. Equation (\ref{7410}) takes the form
\beq
\frac{\d V^k}{\d t}+ V^j \d_j V^k=\frac{\d v^k}{\d t} -D\, \d^k \l( \frac{\d P/ \d t}{P}\r) +(v^j - D \,  \d^j \ln P) \, \d_j (v^k-D \, \d^k \ln P).
\eeq
Accounting for the Fokker-Planck equation it becomes:
\begin{eqnarray}
\frac{\d V^k}{\d t}+ V^j \d_j V^k=\l( \frac{\d v^k}{\d t}+v^j \d_j v^k \r)
 -D\, \d^k \l( \frac{D \Delta P -\d_j P\,  v^j-P \, \d_j v^j}{P} \r)-\nonumber\\ 
 -D\, v^j \d_j \d^k \ln P 
 -D\, \d^j \ln P \,  \d_j v^k 
 +D^2 \, \d^j \ln P \,  \d_j \d^k \ln P.
\end{eqnarray}
After some calculation one finally obtains:
\beq
\frac{\d V^k}{\d t}+ V^j \d_j V^k=- 2 D^2 \, \d^k \l( \frac{\d_j \d^j \sqrt{P}}{\sqrt{P}} \r) + \frac{d v^k}{dt}+D \,\{ \d^k \d_j v^j+ (\d^k v^j-\d^jv^k)\d_j \ln P\}.
\eeq
In the case when $v$ is potential, the last rotational term vanishes and the force in the right-hand side of this equation derives itself from a potential
\beq
\Phi=2 D^2 \, \frac{\Delta \sqrt{P}}{\sqrt{P}}-D \, \Delta \varphi+ \frac{\d \varphi}{\d t} + \frac{1}{2} (\nabla \varphi)^2, 
\eeq
where we have set $v= \nabla \varphi$. This is in particular the case of the scale-relativistic description (see Sec.~\ref{sec2.1}) where $v=v_+$ is potential. The quantum potential (plus possibly an external potential $\phi$) can therefore be obtained in this case provided
\beq
 \frac{\d \varphi}{\d t} + \frac{1}{2} (\nabla \varphi)^2-D \, \Delta \varphi= \phi -4D^2 \, \frac{\Delta \sqrt{P}}{\sqrt{P}}.
\eeq
Under this condition, the Euler and continuity equations can be integrated under the form of a Schr\"odinger equation,
\beq
{D}^2 \Delta \psi + i {D} \frac{\d \psi}{\d t}=\frac{1}{2} \phi \,  \psi.
\eeq
Therefore the possible values of the velocity field $v=v_+$ differ fundamentally  between the two situations (quantum versus diffusion). In particular $v_+=U+V=0$ is excluded in the quantum case, since it leads to the standard diffusion equation. This explains how the Fokker-Planck equation can be common to the two processes, despite their fundamental antinomy.

%¡¡¡¡¡¡¡¡¡¡¡¡¡¡¡¡¡¡¡¡¡¡¡¡¡¡¡¡¡¡¡¡¡¡¡¡¡¡¡¡¡¡¡
\section{Conclusion and future prospect}
%¡¡¡¡¡¡¡¡¡¡¡¡¡¡¡¡¡¡¡¡¡¡¡¡¡¡¡¡¡¡¡¡¡¡¡¡¡¡¡¡¡¡¡
In this paper, the equivalence between the scale relativity geodesic form of the motion equation, its Schr\"odinger form and its hydrodynamical form including a quantum potential has been established in several situations. The Euler and continuity equations with a quantum potential can be always integrated and combined under the form of a linear (without pressure) or non-linear (with pressure) Schr\"odinger equation (NLSE) when the fluid motion is irrotational. 

In the case of fluid motion including vorticity and therefore possibly turbulence, we have also obtained a Schr\"odinger form for a large class of possible flows. Such an equation is of the magnetic Schr\"odinger form, in which the vorticity field plays a role similar to that of an exterior electromagnetic field. Future work is needed to generalize this result to more general flows: we consider the possibility to use more complete tools, such as spinorial, bispinorial or multiplet wave functions (see \cite{MNC08} for a recent attempt of implementation of this proposal), and a generalized description of the vectorial vorticity field, using, e.g., non-Abelian gauge field theory \cite{NCL06}.

The same transformation also holds for a classical charged fluid subjected to an electromagnetic field to which one also applies a potential having the form of a quantum potential. Such a fluid is then described by a Ginzburg-Landau-like equation, and it is therefore expected to have at least some of the properties of a quantum fluid. This particularly interesting case will be specifically studied in more detail in future works, since it could lead to a new kind of macroscopic superconducting-like behavior.

Another generalization that we shall study in the future concerns the full Navier-Stokes equation including viscosity. In the present paper, we have shown that the viscosity term can be incorporated as a phase-dependent nonlinear term in the NLSE. However, as shown in \cite{LN97}, one may combine the viscosity coefficient and the parameter $\cal D$ in terms of a complex parameter, so that the viscosity term can be included in the covariant derivative of the scale relativity approach. One therefore also obtain a Schr\"odinger form for the fluid equations in this case, but in which the wave function needs a new interpretation.

The same method has been applied to the chaotic rotational motion of a solid, and a generalized  form of the quantum potential has been obtained in this case. Finally, we have shown in this paper that, after a change of variable, the diffusion equation can also be given the form of a continuity and Euler system including an additional potential energy. Since this potential is exactly the opposite of a quantum potential, the quantum behavior may be considered, in this context, to be equivalent to a kind of anti-diffusion. Consequences for the inverse problem (of possible partial reversal of the motion of a diffusive fluid) will be considered in future works.

Let us conclude by recalling that some numerical simulations of possible future experimental devices implementing this theoretical description have given encouraging results \cite{NL06}. Such kind of  devices, in which the applied potential depends on the knowledge of some internal measurable properties of the system (such as a density of matter or a probability density) involves a retro-action loop which may be typical of living-like systems \cite{NA07}.\\

{\bf Acknowledgement.} I acknowledge helpul and enlightening discussions with Drs. Thierry Lehner, Marie-No\"elle C\'el\'erier and Charles Auffray during the preparation of this paper.

%%%%%%%%%%%%%%%%%%%%%%%%%%%%%%%%%%%

%*****************

%%%%%%%%%
\end{document}